\documentclass[12pt,a4]{article}

\usepackage{graphicx}%epsfig}%
\usepackage{color}

\newcommand{\rf}[1]{(\ref{#1})}
\newcommand{\be}{\begin{equation}}
\newcommand{\ee}{\end{equation}}
\newcommand{\bea}{\begin{eqnarray}}
\newcommand{\eea}{\end{eqnarray}}

\newcommand{\Lam}{\Lambda}

\renewcommand{\a}{\alpha}

\newcommand{\del}{\delta}

\newcommand{\dg}{\dagger}

\newcommand{\ra}{\right\rangle}

\def\void{}
\def\labelmark{}

\newenvironment{formula}[1]{\def\labelname{#1}
\ifx\void\labelname\def\junk{\begin{displaymath}}
\else\def\junk{\begin{equation}\label{\labelname}}\fi\junk}%
{\ifx\void\labelname\def\junk{\end{displaymath}}
\else\def\junk{\end{equation}}\fi\junk\labelmark\def\labelname{}}

{\ifx\void\labelname\def\junk{\end{array}\end{displaymath}}
\else\def\junk{\end{array}\right.\end{equation}}
\fi\junk\labelmark\def\labelname{}\def\junk{}
}

\newcommand{\beq}{\begin{formula}}
\newcommand{\eeq}{\end{formula}}
\newcommand{\beqv}{\begin{formula}{}}

\vspace{12pt}

\newcommand{\cuum}{|{\rm vac}\rangle}

\newcommand{\hH}{{\hat{H}}}

\newcommand{\ScF}{L}%a				%Scale Factor
\newcommand{\Accel}{B}%{\Delta}	%\textcolor{red}{A}}%{\nu}
\newcommand{\AccelA}{\Accel'}%{\Accel\hspace{-1pt}'}
\newcommand{\AccelB}{\Accel^{\prime\hspace{1pt}2}}%{\Accel\hspace{-1pt}^{\prime\hspace{1pt}2}}
\newcommand{\cc}{\mu} %\Lambda

\begin{document}

\rightline{\today}

\begin{center}
\vspace{24pt}
{ \Large \bf A modified Friedmann equation}

\vspace{24pt}

{\sl J.\ Ambj\o rn}$\,^{a,b}$,
and {\sl Y.\ Watabiki}$\,^{c}$

\vspace{10pt}

{\small

$^a$~The Niels Bohr Institute, Copenhagen University\\
Blegdamsvej 17, DK-2100 Copenhagen \O , Denmark.\\
email: ambjorn@nbi.dk
\vspace{10pt}

$^b$~Institute for Mathematics, Astrophysics and Particle Physics
(IMAPP)\\ Radbaud University Nijmegen, Heyendaalseweg 135, 6525 AJ, \\
Nijmegen, The Netherlands

\vspace{10pt}

$^c$~Tokyo Institute of Technology,\\ 
Dept. of Physics, High Energy Theory Group,\\ 
2-12-1 Oh-okayama, Meguro-ku, Tokyo 152-8551, Japan\\
{email: watabiki@th.phys.titech.ac.jp}

}

\end{center}

\vspace{24pt}

\begin{center}
{\bf Abstract}
\end{center}

We recently formulated a model of the universe based on 
an underlying W3-symmetry. It allows the creation of 
the universe from nothing and the creation of baby universes and
wormholes  for spacetimes of dimension 2, 3, 4, 6 and 10.
 Here we show that the classical large time and large space limit
 of these universes is one of exponential fast expansion without the need
 of a cosmological constant. Under a number of simplifying assumptions 
 our model predicts that w=-1.2 in the case of four-dimensional spacetime. 
 The possibility of obtaining a w-value less than -1
 is linked to the ability of our model to create baby universes 
 and wormholes.

\newpage

\section{Introduction}

According to observations our universe is presently accelerating. 
Such an acceleration can be caused by the presence of a cosmological constant. 
However, 
the precise origin of such a constant is still unclear. 
Also, modifications of Einstein's theory of general relativity 
without an explicit cosmological term 
can provide explanations of the acceleration. 
In this article we propose yet another explanation of the acceleration
which is caused by the production of baby universes. 
%It links 
%the possibility of an acceleration to creation of our present universe. 

In \cite{aw1,aw2} we have formulated a pre-geometric $W_3$-symmetric theory.
The operators of this theory are represented in a large Hilbert space and the 
 ``Hamiltonian'' of this theory has no geometric interpretation and strictly
speaking not even an interpretation as a Hamiltonian since we do not yet have 
a concept of time, and finally there  are no coupling constants associated with this 
algebraic structure which we denote the   ``$W$-algebra world'' (WAW), although admittedly 
there is not much ``world'' at this stage.  However, the WAW can be written as a 
direct sum of Hilbert spaces where  eigenstates of certain of the $\a_k$ operators associated 
with the $W_3$ model, coherent states 
relative to the ``absolute vacuum state'' of the $W_3$ algebra, play the role of physical vacuum states.
Projected onto one of the Hilbert spaces associated with these vacua, the $W_3$ ``Hamiltonian" can 
be given an interpretation as a real Hamiltonian multiplied by a time-coordinate, and 
the expectation values of the $\a_k$ operators associated with the chosen coherent state will define 
the coupling constants of the Hamiltonian. One can view the choice of a coherent state
as a kind of spontaneous symmetry breaking of the $W_3$ symmetry which results in the emergence of time
and a corresponding Hamiltonian. This Hamiltonian has the potential to spontaneously create ``space''
from nothing, to evolve space in time and to merge and split such universes.

%%  where the different Hilbert spaces of $W$-algebra 
%%  interact each other in this world.
%%  We call this world \lq\lq $W$-algebra world (WAW)". 
%%  One of such interactions creates the coherent state 
%%  which causes the emergence of time and a corresponding Hamiltonian, 
%%  and then the successive spontaneous creation and evolution of spaces. 
% The coupling constants of this theory of geometries 
%are defined by spontaneous symmetry breaking in much the same way as 
%some of the coupling constants of the standard model are defined by 
%the spontaneous symmetry breaking of electroweak $SU(2)$ symmetry. 
We will show that under certain assumptions 
the large time behaviour of our model is that of exponential fast expansion precisely as 
in the  standard $\Lam$-DCM model,
but the origin of the acceleration is not a cosmological constant $\Lam$, 
but another parameter we denote $\Accel$, 
related to the $W_3$ symmetry breaking and more specifically to the splitting and merging of 
spatial universes.

\section{The ``classical'' Hamiltonian}

The $W_3$-model studied in \cite{aw1} allows for the  creation of  a two-dimensional 
expanding universe from ``nothing''.  We will first discuss the dynamics of the two-dimensional model 
and then generalize it to four-dimensional spacetime. Let us briefly describe the two-dimensional model
(for details we refer to \cite{aw1,aw2}). A $W_{3}$ algebra\footnote{For discussions of $W_3$ algebras in 
relation to gravity and string theory see \cite{W3}.} is defined in terms of  
operators $\a_n$ satisfying   
\be\label{jx10}
[\a_m,\a_n] = m \,\del_{m+n,0}.
\ee
The so-called $W$-Hamiltonian ${\hH}_{\rm W}$ is then defined by 
\be\label{jy3}
{\hH}_{\rm W} := 
-\,\frac{1}{3} \sum_{k, l, m = -2}
:\!\alpha_k \alpha_l \alpha_m\!: 
\ee
where the normal ordering $:\!\!(\cdot)\!\!:$ 
refers to the $\a_n$ operators ($\a_n$ to the 
left of $\a_m$ for $n>m$). Note that ${\hH}_{\rm W}$ 
does not contain any coupling constants. 
The ``absolute vacuum'' $\left | 0\ra$ is defined by:
\begin{equation}\label{jy2}
\a_n |0\rangle =  0,\quad n  \leq 0.
\end{equation}
By introducing a coherent state, 
which is an eigenstate of $\a_0$, $\a_{-1}$ and $\a_{-3}$ and which 
we denoted the ``physical'' vacuum state $\cuum$, ${\hH}_{\rm W}$ 
is closely related to the so-called generalized causal dynamical triangulation (CDT\footnote{CDT is an attempt to define 
non-perturbatively a theory of fluctuating geometries. 
The theory can be partially solved for  two-dimensional spacetimes
(see \cite{al,GCDT} which also describes the summation over all genera) and it has a string field theoretical formulation
\cite{GSFT}, which is build on the old formulation of non-critical string field theory \cite{NCSFT}. Finally CDT can 
be generalized to higher dimensions, see \cite{physrep} for a review and \cite{higherCDT} for the original literature.}) string field Hamiltonian   $\hH$ 
(see \cite{GSFT} for a definition and details), 
and the coupling constants of $\hH$ are defined via the expectation values
of the operators $\a_0$, $\a_{-1}$ and $\a_{-3}$.
More precisely one has \cite{aw1}
 \be\label{zj2}
 {\hH}_{\rm W}  \propto \hH + c_4 \a_4 +c_2 \a_2
 %+ c_1 \phi_1^\dg + c_0,
 \ee
where $\hH$ is the CDT string field Hamiltonian with standard creation 
and annihilation operators acting on the physical vacuum state $\cuum$, 
and $c_4$ and $c_2$ are nonzero constants. 
This equation shows that the choice of physical vacuum leads to 
the CDT string field Hamiltonian $\hH$ 
except for the term $c_4 \a_4 +c_2 \a_2$ 
which will spontaneously create universes of infinitesimal lengths. 
Let $L$ denote a macroscopic length of a  spatial universe (which we assume has the topology of a circle) 
and let $\Psi^\dg(L)$ be the operator which creates this spatial universe by acting on  $\cuum$. The relation 
between $\Psi^\dg (L)$ and the $\a_n$ are as follows 
\be\label{jj1}
\Psi^\dagger(L) = \sum_{l=0}^\infty \frac{L^l}{l!} \, \a_l,
%,\qquad
%\phi_l = \int_0^\infty \d L \;
%\frac{L^l}{l!} \; \Psi(L)
\ee
and one can now express the string field Hamiltonian $\hH$ 
in terms of $\Psi(L)$ and $\Psi^\dagger(L)$:
\be\label{s8}
\hH = \hH_0 - g \int\!\!\!\int  \!dL_1 dL_2 \;\Psi^\dg(L_1)\Psi^\dg(L_2)\;
(L_1\!+\!L_2)\Psi(L_1\!+\!L_2) + \cdots
%\\ && -  g\int\!\!\!\int \!dL_1 dL_2 \;\Psi^\dg(L_1\!+\!L_2)\;L_2\Psi(L_2)\;L_1\Psi(L_1)
%-\int \!dL \; \delta(L) \Psi(L), \nonumber
\ee
where $\hH_0$ is the original CDT Hamiltonian \cite{al} which does not allow for splitting and 
merging of spatial universes, while the next term describes the splitting of a universe in two (see \cite{GSFT} for 
the terms indicated by $\cdots$ which contain a term for merging two universes to one and a term 
which allows a universe of length zero to vanish). The coupling constant $g$ 
can be expressed in terms of the expectation values of the operators $\a_0$ and $\a_{-3}$. 
The  classical theory based on $\hH$, but  describing only 
the time evolution of a single universe,  has as dynamical 
variables $L$ and its conjugate momentum $\Pi$, 
which have the following Poisson bracket and Hamiltonian
\begin{equation}\label{PoissonBracketLPi}
\{ L \!\;, \Pi \} \,=\, 1,
\end{equation}
\begin{equation}\label{HamiltonianLPi}
{\cal H} = N L \bigg( \Pi^2 - \cc + \frac{2g}{\Pi} \bigg).
\end{equation}
Here $\mu$-term is a cosmological term (which we later will dispose of) 
and $N(t)$ is a Lagrange multiplier field, 
expressing the invariance of the dynamical system under time-reparametrisation.  
The $g$-term in \rf{HamiltonianLPi} comes from 
merging and splitting terms in \rf{s8}
while the value of $\mu$ is the product of $-2g$ 
and the expectation values of $\a_{-1}$ 
(see \cite{aw1} for details). The Hamiltonian \rf{HamiltonianLPi} differs 
from the classical limit of $\hH_0$ \cite{agsw} 
by the presence of the $g$-term
%\textcolor{red}{
which, as mentioned, is related to the merging 
and splitting of spatial universes and thus to the production of baby universes.
It is known that the production of baby universes is intimately related to the 
the fractal structure of spacetime \cite{fractal} and thus 
the presence of the $g$-term originates from a genuine  {\it quantum} gravity effect.
%}

The time dependences of $L$ and $\Pi$ are now given by
\begin{eqnarray}
\dot{L} \!&=&\! -\,\{ {\cal H} \!\;, L \}
\,=\, 2 N L\bigg( \Pi - \frac{g}{\Pi^2} \bigg)
,
\label{dotL}
\\
\dot{\Pi }\!&=&\! -\,\{ {\cal H} \!\;, \Pi \}
\,=\, -\,N \bigg( \Pi^2 - \cc + \frac{2g}{\Pi} \bigg)
.
\label{dotPi}
\end{eqnarray}
We can solve these equations for $\Pi$ by noting that  
\rf{dotL}  can be  written as
\begin{equation}
\frac{\dot{L}}{2 g^{1/3} N L}
\,=\,
\frac{\Pi}{g^{1/3}}
-
\bigg( \frac{g^{1/3}}{\Pi} \bigg)^{\!2}
.
\end{equation}
Thus we find 
\begin{equation}\label{Pi_by_dotL}
\Pi  =  \frac{\dot{L}}{2 N L} F(x)
,
\qquad\quad
x := - \frac{8g N^3 L^3}{\dot{L}^3}
,
\end{equation}
where $F(x)$  satisfies
\begin{equation}\label{ConditionF}
\big( F(x) \big)^3 - \big( F(x) \big)^2 + x
\ = \ 0
.
\end{equation}
The regions of $x$ and $F$ are 
$x \le \frac{4}{27}$
and 
$F \ge \frac{2}{3}$, respectively,
if we start from a universe with zero length. 
From this we find the Lagrangian
\be\label{ActionL}
{\cal L} = 
%\int\! d t \, \big(\!
 - {\cal H} + \Pi \dot{L} =
%\big)
%\nonumber\\&=&\!
%\int\! d t \, \bigg\{\!
  %- N L \bigg( \Pi^2 - \cc + \frac{2g}{\Pi} \bigg)
  % + 2 \Pi N L \bigg( \Pi - \frac{g}{\Pi^2} \bigg)
%\bigg\}
%\nonumber\\&=&\!
%%\int\! d t \, N L \,
%\bigg\{\!
%  - \Pi^2 + \cc - \frac{2g}{\Pi}
%  + 2\!\> \Pi^2 - \frac{2g}{\Pi}
%%\bigg\}
%\nonumber\\&=&\!
%\int\! d t \,
 N L 
 \bigg(
   \Pi^2 + \cc - \frac{4g}{\Pi} \bigg)
 %=  N L\,f\bigg(\frac{\dot{L}}{N L}\bigg)
% =  N L \bigg(
%      \cc
%      + \bigg( \frac{g}{x} \bigg)^{\!2/3}\!
%        F(x) \big( 4 - 3 F(x) \big)
%    \bigg)
,
 \ee
where $\Pi$  is replaced by the right-hand side of 
the first equation of \rf{Pi_by_dotL}. Finally the 
equation which describes the expansion of the universe 
(our modified Friedmann equation without matter)  is 
\begin{equation}\label{FriedmannEqWithoutMatter}
\frac{\delta S}{\delta N} = 0
,
\qquad\quad
S := \int\! d t \!\; {\cal L}.
\end{equation}
One can show that the other  equation of motion,
%\begin{equation}\label{FriedmannEq2WithoutMatter}
$\delta S/\delta L= 0$,
%\end{equation}
is just the time derivative of eq.\ \rf{FriedmannEqWithoutMatter}.

\section{After the Big Bang}

As mentioned above 
the $W_3$-model has
the potential to describe how the universe emerges from a pre-geometric theory 
where there is no interpretation of time and space. 
Space born with zero length expands its size $L(t)$ as \cite{aw2}
\be\label{tht}
L(t) \approx \frac{1}{\sqrt{\mu}} \tanh (\sqrt{\mu} \, t) \qquad {\rm or} \qquad    
L(t) \approx \frac{1}{\sqrt{-\mu}} \tan (\sqrt{-\mu} \, t) 
\ee
depending on the sign of the cosmological constant $\mu$ in \rf{ActionL}.
We denote this the THT-expansion, referring to the functions appearing in \rf{tht}. 
However, this ``classical'' expansion  of space from zero length to a macroscopic size 
(i.e.\ a size larger than or equal the Planck length)  
is complicated by the creation and annihilation of spatial universes 
dictated by the CDT string field Hamiltonian. 
%Potentially this phase can also solve the horizon problem in the same way as the standard  inflation scenario \cite{guth}, 
%but a detailed statement requires a non-perturbative solution of the baby universe dynamics 
%leads to an inflation-like scenario 
%(see \cite{aw2}). 

In this article we will be interested in the other limit, 
that of large time and large space extension 
and a situation where matter fields are included in the model.
In principle one can include matter fields in the string field Hamiltonian, 
but we here take the phenomenological 
approach that for large space extensions 
and large times we insert the classical energy-momentum tensor in the 
classical equations resulting from  \rf{ActionL}. 
Also, we will drop the $\cc$-term in \rf{ActionL} 
which is like a cosmological term. 
We will not need it and in \cite{aw2} we argued 
that effectively it might be zero by quantum effects 
caused by wormholes (the Coleman mechanism \cite{coleman})
when a universe reaches Planck size or larger. 
We will assume that the $g$-term remains in the macroscopic region, 
although quantum effects could change its bare value, 
so we denote the effective macroscopic value $\Accel$. 
Since  $g$ originates from interaction terms in the string field 
Hamiltonian describing the splitting of a spatial universe in two or the merging 
of two spatial universes into one, 
$g$ is also related to the creation of  spacetime wormholes  and is thus  
linked to specific  geometric properties of spacetime. 
Therefore it is difficult to imagine that local dynamics effectively can force it to be zero,
and the assumption of a non-vanishing $g$-tem  in the ``classical'' hamiltonian \rf{ActionL} is  quite natural.
Thus effectively we make the replacement\footnote{Strictly speaking the correct replacement when the spatial
dimension is  $d =2,3,\ldots$ is  $\cc \to \frac{1}{2d (d-1)} \kappa \rho$. Since we will mainly consider the 
case $d=3$ below we have used the substitution  $\cc \to \frac{1}{12} \kappa \rho$ for all $d$ in order
to simplify the notation.} 
$\cc \to \frac{1}{12} \kappa \rho$, $g \to -\frac{1}{8} \Accel$ 
and with these substitutions we obtain 
\begin{equation}\label{FriedmannEqWithMatter}
\frac{\delta S}{\delta N} = 0,
\qquad\quad
S := \int\! d t \!\; {\cal L}
\,\Big |_{\cc \to \frac{1}{12} \kappa \rho,\, g \to  -\frac{1}{8} \Accel}
,
\end{equation}
where 
$\kappa$ is a positive constant which is proportional to the gravitational  constant and 
$\rho$ is the energy density of matter. 
This is our modified Friedmann equation. 
After some algebra it can be written as  
\begin{equation}\label{ModifiedFriedmannEq}
\frac{\dot{L}^2}{N^2L^2}
\,=\,
    \frac{\kappa \rho}{3}
  + \frac{\Accel N L}{\dot{L}}\,
    \frac{1 + 3 F(x)}
         {\big( F(x) \big)^2},
\qquad\quad
x := \frac{\Accel N^3 \ScF^3}{\dot{\ScF}^3}.
\end{equation}
%It has the following expansion in powers of $\Accel$:
%\begin{eqnarray}\label{ModifiedFriedmannEq2D}
%\frac{\dot{L}^2}{4 N^2 L^2}
%\!&=&\!
%\frac{\kappa \rho}{12}
%+ \frac{4 \Accel N L}{\dot{L}}
%+ \frac{20 \Accel^2 N^4 L^4}{\dot{L}^4}
%+ O(\Accel^3)
%\end{eqnarray}
Making the gauge choice $N(t) =1$ 
we see that under the assumption that $\rho \to 0$ for $L \to \infty$
we have for large $t$ that $\dot{L}/L$ is constant, 
and we find
\begin{equation}
\frac{\dot{L}}{L} \to 3 \sqrt[3]{\Accel/4},\qquad {\rm i.e.} \qquad 
L \,\sim\, e^{\,3 \sqrt[3]{\Accel/4}\, t}.
\end{equation}
This value is also the boundary value $x = \frac{4}{27}$,
so the Hubble parameter $H := \dot{L} / L$ takes values
in the region $3 \sqrt[3]{\Accel/4} \le H \le \infty$.
Thus the $\Accel$-term acts effectively 
as a cosmological term for large $t$ in the sense that 
it can lead to an exponentially expanding universe for large t.

%\section{The four-dimensional modified Friedmann equation}

The two-dimensional $W_3$ symmetry breaking scenario was 
generalized to 3, 4, 6 and 10 dimensional spacetime 
by considering $W_3$ algebras with intrinsic symmetries \cite{aw2}. 
Like the ordinary Friedmann equation, which assumes 
homogeneity and isotropy of space 
and which is independent of the dimension of space, 
we expect that our modified Friedmann equation 
will also be independent of the dimension of space 
if we make the same assumptions.
%\footnote{Strictly speaking the $\rho$-term has 
%some $d$-dependence. The natural substitution would be $\mu \to \kappa \rho / (2d(d-1))$ where
%$d$ is the spatial dimension. Below we will use it for $d=3$. That is the reason we used the specific 
%factor 1/12 in \rf{FriedmannEqWithMatter}-\rf{ModifiedFriedmannEq} such that they agree with the 
%standard equations for $d=3$.}.
Here we will concentrate on the four-dimensional spacetime case 
and thus write the modified Friedmann equation \rf{ModifiedFriedmannEq} 
(identifying the conventional scale factor $a(t)$ 
with $L(t)$ and gauge fixing to  $N(t) = 1$).
 \begin{equation}\label{ModifiedFriedmannEqA}
%\bigg(
 \frac{\dot{\ScF}^2}{\ScF^2}
%\bigg)^{\!2}
\,=\,
\frac{\kappa \rho}{3}
  + \frac{\Accel \ScF}{\dot{\ScF}}\,
    \frac{1 + 3 F(x)}
         {\big( F(x) \big)^2}
%+ \frac{\AccelA \ScF}{3 \dot{\ScF}}
%+ O(\AccelB)
,
\qquad\quad\hbox{where}\quad
x := \frac{\Accel \ScF^3}{\dot{\ScF}^3}.
\end{equation}
%In this equation $\Accel$ is the four-dimensional parameter 
%corresponding to our to $\nu$ in the two-dimensional case. 
One quantity in \rf{ModifiedFriedmannEqA} {\it will} 
depend on the dimension of space,
namely $\rho$. For simplicity we will here assume 
we have a simple matter dust system. Thus the pressure of matter
is zero and for our three-dimensional space we have 
\be\label{ja2}
\kappa \rho%(a,t)
 = \frac{C}{\ScF^3},
\quad 
\dot{\rho} = -\, 3 H \rho,
\quad 
H(t) := \frac{\dot{\ScF}}{\ScF},
\ee
where $H(t)$ is the usual (time dependent) Hubble parameter
and $C$ is a positive constant. 
On the right-hand side of equation \rf{ModifiedFriedmannEqA}, 
the first term is the matter energy and 
the second term is a new type of term
(we have chosen the cosmological constant $\Lam =0$ so we do not 
have a ``standard" dark energy term).
In order to make clearer the ratio of these ``energies'',
\rf{ModifiedFriedmannEqA} can be written as
\begin{equation}\label{ModifiedFriedmannEqAratio}
\Omega_{\rm m}(t)
+
\Omega_\Accel(t)
+
\Omega_K(t)
+
\Omega_\Lambda(t)
\,=\,
1
,
\end{equation}
where
\begin{equation}\label{ModifiedFriedmannEqOmega}
\Omega_{\rm m}(t)
\,:=\,
\frac{\kappa \rho}{3}
\bigg( \frac{\ScF}{\dot{\ScF}} \bigg)^{\!2}
,
\qquad
\Omega_\Accel(t)
\,:=\,
\frac{x \big( 1 + 3 F(x) \big)}
     {\big( F(x) \big)^2}
,
%\frac{\AccelA}{3}
%\bigg( \frac{\ScF}{\dot{\ScF}} \bigg)^{\!3},
%+ O(\AccelB)
\qquad
\Omega_K(t) = \Omega_\Lambda(t) = 0
.
\end{equation}
In our model the topology of space is toroidal, 
so the space is flat, i.e.\ $\Omega_K(t) = 0$.
Moreover,
the cosmological constant is assumed to be zero 
because of the Coleman mechanism,
so the dark energy is also zero, i.e.\ $\Omega_\Lambda(t) = 0$.
%\begin{equation}
%\Omega_K = \Omega_\Lambda
%\end{equation}
We now write the modified Friedmann equation as 
\begin{equation}\label{ModifiedFriedmannEqALambdaZero2}
\frac{\dot{\ScF}^2}{\ScF^2}
\,=\,
\frac{C}{3 \ScF^3}
+ \frac{4 \Accel \ScF}{\dot{\ScF}}
+ \frac{5 \Accel^2 \ScF^4}{\dot{\ScF}^4}
+ \ldots%O(\Accel^2)
\end{equation}
and solve it perturbatively in $\Accel$, 
starting with the boundary condition\footnote{$L(0) =0$ is only intended to mean that $L(t)$ is small, but 
macroscopic for small $t$. For the detailed description of  universes of Planck scales one has to 
use the full CDT string field Hamiltonian.} that $\ScF(t) = 0$ for $t=0$:
\be\label{MatterDominantEq3}
\ScF(t)  \,=\,
  \bigg( \frac{3 C}{4} \bigg)^{\!1/3} \bigg(
    t^{2/3}
    + \frac{9\!\> \Accel\!\> t^{11/3}}{8}
    - \frac{1539\!\> \Accel^2\!\> t^{20/3}}{1792}
    + \ldots%O(\Accel^2)
  \bigg)
.
\ee
%Let us eliminate $B'$ in favour of the starting time of acceleration, 
%\begin{eqnarray*}
%\AccelA
%\,=\,
%\frac{32}{135 \!\;t_\ast^3}
%+ O(\AccelB)
%\,=\,
%\frac{32}{135 \!\;t_\ast^3}
%+ O\bigg(\frac{1}{t_\ast^6}\bigg).
%\end{eqnarray*}
%Then we can write
%\be\label{MatterDominantEq4}
%\ScF \,=\,
%  \alpha \bigg(
%    t^{2/3}
%    + \frac{t^{11/3}}{45 \!\; t_\ast^3}
%    + O\bigg(\frac{1}{t_\ast^6}\bigg)\\
%  \bigg)
%\ee
%and the Hubble parameters become
%\begin{eqnarray}
%\frac{\dot{\ScF}}{\ScF}
%\!&=&\!
%\frac{2}{3 \!\> t} \bigg(
%  1 + \frac{t^3}{10 \!\; t_\ast^3}
%  + O\bigg(\frac{1}{t_\ast^6}\bigg)
%\bigg)
%\nonumber\\&=&\!
%\frac{2}{3 \!\> t}
%+ \frac{3 \!\; t^2}{44 \!\; t_\ast^3}
%+ O\bigg(\frac{t^5}{t_\ast^6}\bigg)
%\label{HubbleParameter2}
%\\
%\frac{\ddot{\ScF}}{\ScF}
%\!&=&\!
%-\, \frac{2}{9 \!\> t^2} \bigg(
%  1 - \frac{t^3}{t_\ast^3}
%  + O\bigg(\frac{1}{t_\ast^6}\bigg)
%\bigg)
%\nonumber\\&=&\!
%-\, \frac{2}{9 \!\> t}
%+ \frac{5 \!\; t^2}{22 \!\; t_\ast^3}
%+ O\bigg(\frac{t^5}{t_\ast^6}\bigg)
%\label{RedShiftParameter2}
%\\
%-
%\bigg( \frac{\ScF}{\dot{\ScF}} \bigg)^{\!2}
%\frac{\ddot{\ScF}}{\ScF}
%\!&=&\!
%\frac{1}{2} \bigg(
%  1
%  - \frac{6 \!\; t^3}{5 \!\;t_\ast^3}
%  + O\bigg(\frac{1}{t_\ast^6}\bigg)
%\bigg)
%\label{DecelerationParameter2}
%\end{eqnarray}
When the time $t$ increases from $0$ to $\infty$,
$x$ increases from $0$ to $\frac{4}{27}$
and
$F$ decreases from $1$ to $\frac{2}{3}$.

\begin{figure}[t]{
\centerline{\scalebox{1.0}{\rotatebox{0}{\includegraphics{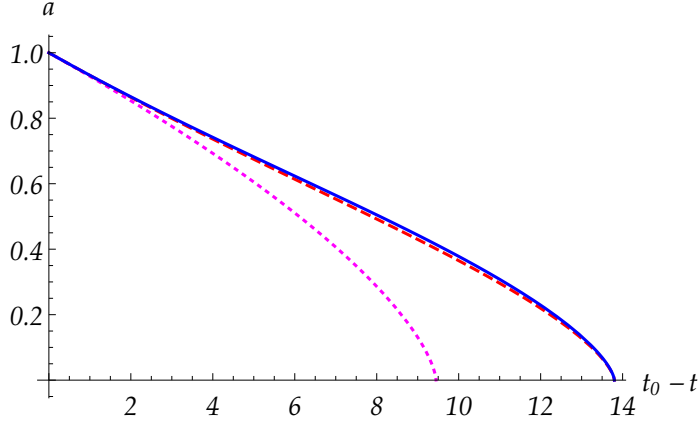}}}}
\caption[fig1]{{\small
The solid line, the dashed line and the dotted line 
are 
$a(t) := \frac{L(t)}{L(t_0)}$ of our model, 
the $\Lambda$-CDM model and 
the CDM model, respectively}}
\label{figScaleFactor}
}
\end{figure}

Lastly, 
we use as observed parameters $t_0$, the present age of the universe, 
and $H_0 := H(t_0)$, the present value of the Hubble parameter. 
In the standard CDM model $H_0$ 
determines the present age of the universe $t_0$. 
Our model needs one more experimental input besides the present Hubble parameter $H_0$
because there exists one more parameter $\Accel$. 
We here use the present time $t_0$ 
in order to fix the parameter $\Accel$. 
We choose
\begin{equation}\label{InitialParameters_tH}
t_0
=
13.8
\,[\,{\rm Gyr}\,]
,
\qquad\quad
H_0 %= \frac{\dot{\ScF}}{\ScF}
=
%73
%72
%71
69
%67.15
\,[\,{\rm km}\,{\rm s}^{-1}\!\;{\rm Mpc}^{-1}\,]
.
\end{equation}
These values of $t_0$ and $H_0$ are based on the recent observations
%the age of the oldest star \cite{xxx}.
%[\,Cf. arXiv:1302.3180v1\,]
and determine the function $\ScF(t)$ \rf{MatterDominantEq3} 
without ambiguity except for the factor $C$. 
The graph of $a(t) := L(t) / L(t_0)$ together with 
those of the $a(t)$'s of the CDM model and the $\Lambda$-CDM model 
are shown in Fig.\,\ref{figScaleFactor}.
One finds that 
%the $\ScF(t)$ of our model is very similar to 
%the $\ScF(t)$ of $\Lambda$-CDM model.
the expansion of the universe {\it as well as the late acceleration} in our model 
is very similar to that in the $\Lambda$-CDM model.
%and is accelerating as the same as the $\Lambda$-CDM model. 

\begin{figure}{
\centerline{\scalebox{1.0}{\rotatebox{0}{\includegraphics{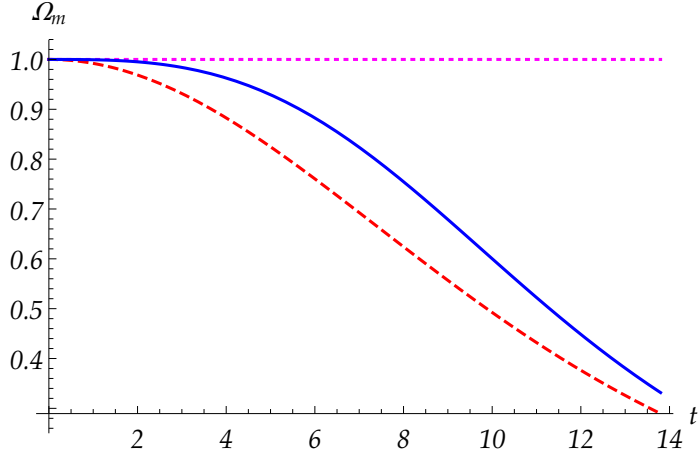}}}}
\caption[fig2]{{\small
The solid line, the dashed line and the dotted line 
are 
$\Omega_{\rm m}(t)$ of our model, 
the $\Lambda$-CDM model and 
the CDM model, respectively}}
\label{figOmega}
}
\end{figure}

\begin{figure}{
\centerline{\scalebox{1.0}{\rotatebox{0}{\includegraphics{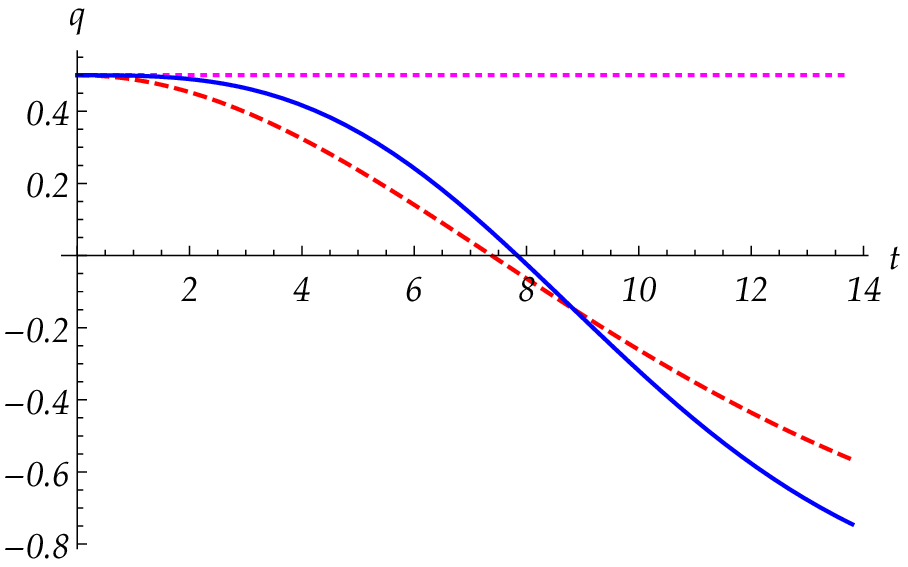}}}}
\caption[fig3]{{\small
The solid line, the dashed line and the dotted line 
are 
$q(t)$ of our model, 
the $\Lambda$-CDM model and 
the CDM model, respectively}}
\label{figQ}
}
\end{figure}

\begin{figure}{
\centerline{\scalebox{1.0}{\rotatebox{0}{\includegraphics{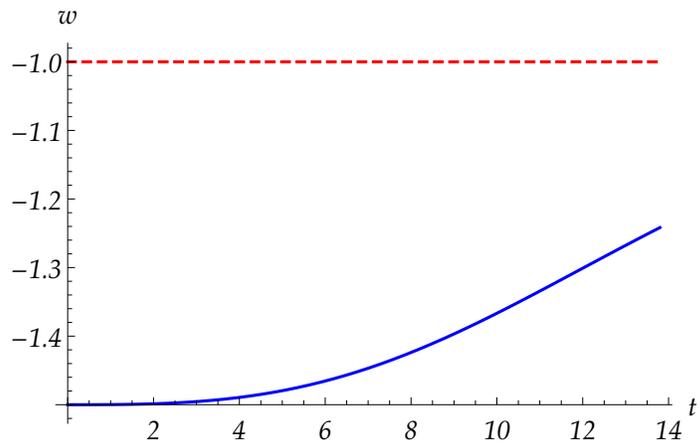}}}}
\caption[fig3]{{\small
The solid line and the dashed line 
are 
$w(t)$ of our model and 
the $\Lambda$-CDM model, respectively}}
\label{figW}
}
\end{figure}

Since $\ScF(t)$'s of our model and the $\Lam$-CDM model are very similar, 
we need to distinguish them carefully.
Using %$t_0$ and $H_0$ of \rf{InitialParameters_tH}, besides 
$\ScF(t)$,
we obtain the 
$\Omega_{\rm m}(t)$, $q(t)$ and $w(t)$ of our model, defined by 
%the deceleration parameter
%$- \frac{\ScF \ddot{\ScF}}{\dot{\ScF}^2}$
\begin{eqnarray}
\Omega_{\rm m}(t)
\,:=\,
\frac{\kappa \rho}{3}
\bigg( \frac{\ScF}{\dot{\ScF}} \bigg)^{\!2},
\qquad
q(t) :=
-
\bigg( \frac{\ScF}{\dot{\ScF}} \bigg)^{\!2}
\frac{\ddot{\ScF}}{\ScF},
\qquad
w(t) :=
-\,\frac{1 - 2 q}{3 \!\> \Omega_\Accel},
%=
%-\,\frac{2 \frac{\ddot{L}}{L} + \big( \frac{\dot{L}}{L} \big)^2}
%        {3 \big( \frac{\dot{L}}{L} \big)^2 - \kappa \rho},
\end{eqnarray}
respectively, 
and we show the graphs in Figs.\,\ref{figOmega}--\ref{figW}.
The $q(t)$ and $w(t)$ of our model are 
different from $q(t)$ and $w(t)$ of $\Lambda$-CDM model.

Define
$\Omega_{{\rm m}0}:= \Omega_{\rm m}(t_0)$, 
$q_0 := q(t_0)$ and 
$w_0 := w(t_0)$, as well as the time $t_\ast$ where the acceleration of the universe is zero, 
%is the values of 
%$\Omega_{{\rm m}0}$, $q_0$ and $w_0$ 
%which are 
%$\Omega_{\rm m}(t)$, $q(t)$ and $w(t)$ at $t_0$, respectively. 
defined by $\ddot{\ScF}(t_\ast) \!=\! 0$, 
and the corresponding value of the red shift  
%$z_\ast$ ($z_\ast$ is defined by
$z_\ast := \big( \ScF(t_0) - \ScF(t_\ast) \big) /\ScF(t_\ast)$. %)
We then obtain
%the value of $z$ at $t_\ast$).
\begin{eqnarray}\label{ja3}
&&
\Omega_{{\rm m}0} \sim 0.33
,
%
%\quad
%\Omega_\Accel \sim 0.66,
%
\qquad
q_0
\sim 
-0.74
,
\qquad
w_0 \sim -1.2
,
\\
&&
t_\ast \sim
7.8
%8.25\cdots
%8.606\cdots
\,[\,{\rm Gyr}\,]
,
\qquad
z_\ast
%= \frac{\ScF(t_0)}{\ScF(t_\ast)} - 1
\sim
%0.25\sdots
%0.57
0.60
%0.5\cdots
%0.465\cdots
. \label{ja33}
\end{eqnarray}
It is worth noting that the values quoted in \rf{ja3} and \rf{ja33} are not very sensitive to the precise values of 
$H_0$ and $t_0$ cited in \rf{InitialParameters_tH}.
The value of $\Omega_{{\rm m}0}$ of our model 
is somewhat bigger than the corresponding 
$\Omega_{{\rm m}0}$  ($\sim 0.29$) of the $\Lambda$-CDM model. The major difference 
compared to the $\Lambda$-CDM model is that we have a $w_0$ value which is less than
-1. In the conventional scenarios this is usually only possible if one introduces some kind 
of ghost fields. However, that is not necessary in our model.

%because the value of $\Omega_{{\rm m}0}$ in $\Lambda$-CDM model 
%is about $0.29$. 

\section{Discussion}

We have presented an alternative Friedmann equation 
where the solution fits the data as well as the ordinary Friedmann equation, 
but which does not require a cosmological constant 
in order to produce  the presently observed acceleration. 
Of course various $f(R)$ models of gravity 
can also lead to acceleration after a matter dominated deceleration 
and if we only keep the linear term in $B$ in our expansion, 
one will not be able to distinguish our model from a suitable $f(R)$ model 
(see \cite{sotiriou} for a review of $f(R)$ models). 
While the deviation of $f(R)$ from the pure Einstein term 
presently is  mainly dictated by the desire  to match observations, 
the origin of the $B$-term in our model  is ``natural'' 
in the sense that it appears in the $W_3$ models 
for the creation of our universe that 
we have suggested elsewhere \cite{aw1,aw2}.

In this paper we have only dealt with the later times of the universe. 
We used $a(t) =0$ for $t=0$. 
But this $t=0$ was only intended to represent a ``short'' time 
after the Big Bang and we have made no attempt to 
follow the detailed history of the universe related a radiation area 
and a matter dominated area. The reason is 
that our additional term is not important at early times. 
However, as reported in \cite{aw1,aw2} our model 
describes the expansion of the universe from ``absolute'' zero size 
and it has the potential to produce 
an extended universe in a short time scale. 
The reason we do not present here a continuous time-picture 
from ``absolute'' zero time  
to the present time is that 
a non-perturbative treatment is needed to understand the full emergence 
of extended spacetime with a metric 
from the pre-geometric state of the universe. 
In this article we assumed that such an extended spacetime have been formed, 
and we started 
from there and studied the effect of the new term present 
in the evolution equations for late times. 
The  obvious next step to be addressed is to understand 
how the emergence of spacetime from pre-geometry can 
be matched with the standard inflation picture.
Presently our model has nothing to say about the naturalness of the value of 
$\sqrt[3]{\Accel} \sim 0.49 / t_0$. 
%$\Accel \sim 0.12 / t_0^3$. 
Like the cosmological constant $\Lam$ in the $\Lam$-CDM model 
it is mysteriously small compared to the Planck scale. 
However, since we have not yet connected 
the sub-Planckian region of our model 
with the late time region of the model, 
it is possible that the magnitude of $\Accel$ 
has a dynamical origin.

Our model clearly allows for more drastic scenarios than 
just an accelerating universe. It  is born with the ability to 
create baby universes and wormholes, processes which might 
not only be important when the universe is of Planck size.
Clearly it will be of utmost importance to understand how 
such processes manifest themselves when we already 
have created a macroscopic universe like the one 
we inhabit today. If we look at the string field equation 
for splitting and joining spatial universes and naively assume that it can
be applied to three-dimensional spaces, the dynamics of this splitting being 
driven by the effective coupling constant $B$, then the small value of 
$B$ that we seemingly observe would indicate that there is only a small chance of splitting our present universe
in two.  In fact the probability per unit time is more or less equal to the inverse age of our universe. However,
since $B$ is only an effective coupling constant, the dynamics could be very different when the universe at Planck 
time was at Planck size, as already mentioned.

Finally, the selection of a coherent state serving as a physical vacuum,
which has been applied to the WAW world, probably has 
the same features as other more standard scenarios of spontaneous 
symmetry breaking, e.g.\ the symmetry breaking of the Standard Model.
In the Standard Model the vacuum is a coherent state where the Higgs field
has an expectation value different from zero. However, this state may only be 
meta-stable and if that is the case it can decay to another state where the expectation value of 
the Higgs field is zero. This would induce an abrupt change in the coupling constants of 
the Standard Model. Similarly, the coherent state corresponding to one  physical vacuum in 
the WAW could change to another coherent state corresponding to another physical vacuum, 
thereby inducing an abrupt change in the values of the coupling constants of the string field Hamiltonian.
Clearly, these important questions have to be addressed in order understand not only the model, but also
its predictive power.

%Our model also predicts not only the accelerating of universe but also the splitting of universe in the same time scale
%because the coupling constant $g$ appears not only in \rf{HamiltonianLPi} but also in the term of three-universe interaction. 
%This means that our universe will split in near future or might have already splitted recently. Lastly, 
%we comment on the interacting world of different Hilbert spaces of $W$-algebras, i.e.\ WAW. 
%In our previous paper \cite{aw1,aw2}, we point out that the time was born by the interaction 
%between different Hilbert spaces of $W$-algebras,This interaction may cause the sudden change of the value of $g$.
%If this event occured in the past after the Big Bang, the prediction power of our model become weak, but it is possible.

\vspace{24pt} 

\noindent {\bf Acknowledgments.} 

JA and YW acknowledge support from the ERC-Advance grant 291092,
``Exploring the Quantum Universe'' (EQU). 
J.A. acknowledge financial support by the Independent Research Fund Denmark through the grant 
``Quantum Geometry''.


\begin{thebibliography}{99}

 %\cite{Ambjorn:2015spa}
\bibitem{aw1}
  J.\ Ambjorn and Y.\ Watabiki,
  %``A model for emergence of space and time,''
  Phys.\ Lett.\ B {\bf 749} (2015) 149,
  [arXiv:1505.04353 [hep-th]].
  %%CITATION = doi:10.1016/j.physletb.2015.07.067;%%
%\cite{Ambjorn:2017hiw}
\bibitem{aw2}
  J.\ Ambjorn and Y.\ Watabiki,
  %``Creating 3, 4, 6 and 10-dimensional spacetime from W3 symmetry,''
  Phys.\ Lett.\ B {\bf 770} (2017) 252,
  [arXiv:1703.04402 [hep-th]].\\
  %%CITATION = doi:10.1016/j.physletb.2017.04.051;%%
  %1 citations counted in INSPIRE as of 29 Aug 2017
%\cite{Ambjorn:2017dmw}
%\bibitem{Ambjorn:2017dmw}
  %J.~Ambjørn and Y.~Watabiki,
  %``CDT and the Big Bang,''
  Acta Phys.\ Polon.\ Supp.\  {\bf 10} (2017) 299,
  [arXiv:1704.02905 [hep-th]].
  %%CITATION = doi:10.5506/APhysPolBSupp.10.299;%%


%\cite{}
\bibitem{W3}
  L.J.\ Romans,
  %``Realizations of classical and quantum W(3) symmetry,''
  Nucl.\ Phys.\ B {\bf 352} (1991) 829.\\
  %%CITATION = doi:10.1016/0550-3213(91)90108-A;%%
  %106 citations counted in INSPIRE as of 20 Feb 2017
%\bibitem{quantumW}
%\cite{Mohammedi:1991ep}
%\bibitem{Mohammedi:1991ep}
  N.\ Mohammedi,
  %``On gauging and realizing classical and quantum W(3) symmetry,''
  Mod.\ Phys.\ Lett.\ A {\bf 6} (1991) 2977.\\
  %%CITATION = doi:10.1142/S0217732391003481;%%
  %5 citations counted in INSPIRE as of 20 Feb 2017
   %\cite{Figueroa-OFarrill:1994vnq}
%\bibitem{Figueroa-OFarrill:1994vnq}
  J.M.\ Figueroa-O'Farrill,
  %``A Comment on the magical realizations of W(3),''
  Phys.\ Lett.\ B {\bf 326} (1994) 89,
  [hep-th/9401108].\\
  %%CITATION = doi:10.1016/0370-2693(94)91197-5;%%
  %3 citations counted in INSPIRE as of 20 Feb 2017
  %\cite{Lyakhovich:1994xw}
%\bibitem{Lyakhovich:1994xw}
  S.A.\ Lyakhovich and A.A.\ Sharapov,
  %``Quantum inconsistency of W(3) gravity models associated with magical Jordan algebras,''
  hep-th/9411167.\\
  %%CITATION = HEP-TH/9411167;%%
%\cite{Hull:1993kf}
%\bibitem{hull}
  C.M.\ Hull,
  %``Lectures on W gravity, W geometry and W strings,''
  In *Trieste 1992, Proceedings, High energy physics and cosmology* 76-142 and London Queen Mary and Westfield Coll. - QMW-93-02 (93/02) 54 p. (305933)
  [hep-th/9302110].
  %%CITATION = HEP-TH/9302110;%%
  %24 citations counted in INSPIRE as of 20 Feb 2017
 
 
 \bibitem{al}
J.\ Ambjorn and R.\ Loll,
%{\it Nonperturbative Lorentzian quantum gravity, 
%causality and topology change,} 
Nucl.Phys.B {\bf 536} (1998) 407-434, [hep-th/9805108].

  
%\cite{Ambjorn:2007jm}
\bibitem{GCDT}
  J.\ Ambjorn, R.\ Loll, W.\ Westra and S.\ Zohren,
  %``Putting a cap on causality violations in CDT,''
  JHEP {\bf 0712} (2007) 017
  [arXiv:0709.2784 [gr-qc]].\\
  %%CITATION = ARXIV:0709.2784;%%
  %33 citations counted in INSPIRE as of 21 Apr 2015
%\cite{Ambjorn:2008jf}
%\bibitem{Ambjorn:2008jf}
  J.~Ambjorn, R.~Loll, Y.~Watabiki, W.~Westra and S.~Zohren,
  %``A Matrix Model for 2D Quantum Gravity defined by Causal Dynamical Triangulations,''
  Phys.\ Lett.\ B {\bf 665} (2008) 252
  [arXiv:0804.0252 [hep-th]].\\
  %%CITATION = ARXIV:0804.0252;%%
  %34 citations counted in INSPIRE as of 21 Apr 2015
   %\cite{Ambjorn:2009fm}
%\bibitem{sumgenus}
  J.~Ambjorn, R.~Loll, W.~Westra and S.~Zohren,
  %``Summing over all Topologies in CDT String Field Theory,''
  Phys.\ Lett.\ B {\bf 678} (2009) 227
  [arXiv:0905.2108 [hep-th]].\\
  %%CITATION = ARXIV:0905.2108;%%
  %11 citations counted in INSPIRE as of 21 Apr 2015
%\cite{Ambjorn:2013csx}
%\bibitem{Ambjorn:2013csx}
  J.~Ambjorn and T.~G.~Budd,
  %``Trees and spatial topology change in CDT,''
  J.\  Phys.\  A: Math.\  Theor.\  {\bf 46} (2013) 315201
  [arXiv:1302.1763 [hep-th]].
  %%CITATION = ARXIV:1302.1763;%%
  %7 citations counted in INSPIRE as of 21 Apr 2015


%\cite{Ambjorn:2008ta}
\bibitem{GSFT}
  J.~Ambjorn, R.~Loll, Y.~Watabiki, W.~Westra and S.~Zohren,
  %``A String Field Theory based on Causal Dynamical Triangulations,''
  JHEP {\bf 0805} (2008) 032
  [arXiv:0802.0719 [hep-th]].
  %%CITATION = ARXIV:0802.0719;%%
  %38 citations counted in INSPIRE as of 21 Apr 2015




%\bibitem{gs} Giddings and Strominger

%\cite{Ishibashi:1993nq}
\bibitem{NCSFT}
  N.~Ishibashi and H.~Kawai,
  %``String field theory of noncritical strings,''
  Phys.\ Lett.\ B {\bf 314} (1993) 190
  [hep-th/9307045].
  %%CITATION = HEP-TH/9307045;%%
  %82 citations counted in INSPIRE as of 21 Apr 2015
%\cite{Ishibashi:1993nqz}
%\bibitem{Ishibashi:1993nqz}
%  N.~Ishibashi and H.~Kawai,
  %``String field theory of c <= 1 noncritical strings,''
  Phys.\ Lett.\ B {\bf 322} (1994) 67
  [hep-th/9312047].
  %%CITATION = HEP-TH/9312047;%%
  %72 citations counted in INSPIRE as of 21 Apr 2015
%\cite{Ishibashi:1995np}
%\bibitem{Ishibashi:1995np}
%  N.~Ishibashi and H.~Kawai,
  %``A Background independent formulation of noncritical string theory,''
  Phys.\ Lett.\ B {\bf 352} (1995) 75
  [hep-th/9503134].\\
  %%CITATION = HEP-TH/9503134;%%
  %29 citations counted in INSPIRE as of 21 Apr 2015
%\cite{Watabiki:1993ym}
%\bibitem{watabiki}
  Y.~Watabiki,
  %``Construction of noncritical string field theory by transfer matrix formalism in dynamical triangulation,''
  Nucl.\ Phys.\ B {\bf 441} (1995) 119
  [hep-th/9401096].
  %%CITATION = HEP-TH/9401096;%%
  %61 citations counted in INSPIRE as of 21 Apr 2015
%\cite{Watabiki:1994pf}
%\bibitem{Watabiki:1994pf}
%  Y.~Watabiki,
  %``Noncritical string field theory with nonorientable string interactions,''
  Phys.\ Lett.\ B {\bf 346} (1995) 46
  [hep-th/9407058].\\
  %%CITATION = HEP-TH/9407058;%%
  %9 citations counted in INSPIRE as of 21 Apr 2015
%\cite{Ambjorn:1996ne}
%\bibitem{aw}
  J.~Ambjorn and Y.~Watabiki,
  %``Noncritical string field theory for two-D quantum gravity coupled to (p, q) conformal fields,''
  Int.\ J.\ Mod.\ Phys.\ A {\bf 12} (1997) 4257
  [hep-th/9604067].
  %%CITATION = HEP-TH/9604067;%%
  %15 citations counted in INSPIRE as of 21 Apr 2015



\bibitem{physrep}
  J.~Ambjorn, A.~Goerlich, J.~Jurkiewicz and R.~Loll,
% {\it Nonperturbative Quantum Gravity,}
  Phys.\ Rept.\  {\bf 519} (2012) 127
  [arXiv:1203.3591 [hep-th]].
  %%CITATION = ARXIV:1203.3591;%%
  %65 citations counted in INSPIRE as of 06 Dec 2014


\bibitem{higherCDT}
J.~Ambjorn, A.~G\"orlich, J.~Jurkiewicz and R.~Loll,
%{\it Planckian birth of the quantum de Sitter universe},
Phys.\ Rev.\ Lett.\ {\bf 100} (2008) 091304 [arXiv:0712.2485, hep-th].
%%CITATION = PRLTA,100,091304;%%
%\bibitem{bigs4}
%J.~Ambjorn, A.~G\"orlich, J.~Jurkiewicz and R.~Loll,
%{\it The nonperturbative quantum de Sitter universe},
Phys.\ Rev.\  D {\bf 78} (2008) 063544 [arXiv:0807.4481, hep-th].
%%CITATION = PHRVA,D78,063544;%%
%\bibitem{rounds4}
%J.~Ambjorn, A.~G\"orlich, J.~Jurkiewicz and R.~Loll,
%{\it Geometry of the quantum universe},
Phys.\ Lett.\ B {\bf 690} (2010) 420-426 [arXiv:1001.4581, hep-th].\\
%%CITATION = ARXIV:1001.4581;%%
%\bibitem{correlator}
J.~Ambjorn, J.~Jurkiewicz and R.~Loll,
%{\it Emergence of a 4D world from causal quantum gravity},
Phys.\ Rev.\ Lett.\ {\bf 93} (2004) 131301 [hep-th/0404156].
%%CITATION = HEP-TH 0404156;%%
%\bibitem{blp}
%J.\ Ambjorn, J.\ Jurkiewicz and R.\ Loll, 
%{\it Reconstructing the universe},
Phys.\ Rev.\ D\ {\bf 72} (2005) 064014  [hep-th/0505154].
%%CITATION = HEP-TH 0404156;%%
%J.~Ambjorn, J.~Jurkiewicz and R.~Loll,
%{\it Semiclassical universe from first principles},
Phys.\ Lett.\ B {\bf 607} (2005) 205-213
[hep-th/0411152].\\
%%CITATION = HEP-TH 0411152;%%
%\cite{Ambjorn:2010hu}
%\bibitem{Ambjorn:2010hu}
  J.~Ambjorn, A.~Gorlich, S.~Jordan, J.~Jurkiewicz and R.~Loll,
% {\it CDT meets Horava-Lifshitz gravity,}
  Phys.\ Lett.\ B {\bf 690} (2010) 413
  [arXiv:1002.3298 [hep-th]].\\
  %%CITATION = ARXIV:1002.3298;%%
  %53 citations counted in INSPIRE as of 06 Dec 2014
%\cite{Ambjorn:2011cg}
%\bibitem{second-order}
 J.~Ambjorn, S.~Jordan, J.~Jurkiewicz and R.~Loll,
%{\it A Second-order phase transition in CDT,}
Phys.\ Rev.\ Lett.\  {\bf 107} (2011) 211303
[arXiv:1108.3932 [hep-th]].
  %%CITATION = ARXIV:1108.3932;%%
%\cite{Ambjorn:2012ij}
%\bibitem{Ambjorn:2012ij}
%  J.~Ambjorn, S.~Jordan, J.~Jurkiewicz and R.~Loll,
  %``Second- and First-Order Phase Transitions in CDT,''
  Phys.\ Rev.\ D {\bf 85} (2012) 124044
  [arXiv:1205.1229 [hep-th]].
  %%CITATION = ARXIV:1205.1229;%%
  %22 citations counted in INSPIRE as of 06 Dec 2014
%\cite{Ambjorn:2012jv}

\bibitem{agsw}
  J.~Ambjorn, L.~Glaser, Y.~Sato and Y.~Watabiki,
  %``2d CDT is 2d Horava-Lifshitz quantum gravity,''
  Phys.\ Lett.\ B {\bf 722} (2013) 172
  [arXiv:1302.6359 [hep-th]].
  %%CITATION = ARXIV:1302.6359;%%
  %15 citations counted in INSPIRE as of 06 Dec 2014

  
  %\cite{Ambjorn:1995dg}
\bibitem{fractal}
%\cite{Kawai:1993cj}
%\bibitem{Kawai:1993cj}
  H.~Kawai, N.~Kawamoto, T.~Mogami and Y.~Watabiki,
  %``Transfer matrix formalism for two-dimensional quantum gravity and fractal structures of space-time,''
  Phys.\ Lett.\ B {\bf 306} (1993) 19,
  %doi:10.1016/0370-2693(93)91131-6
  [hep-th/9302133].\\
  %%CITATION = doi:10.1016/0370-2693(93)91131-6;%%
  %170 citations counted in INSPIRE as of 24 Oct 2017
  J.~Ambjorn and Y.~Watabiki,
  %``Scaling in quantum gravity,''
  Nucl.\ Phys.\ B {\bf 445} (1995) 129,
 % doi:10.1016/0550-3213(95)00154-K
  [hep-th/9501049].\\
  %%CITATION = doi:10.1016/0550-3213(95)00154-K;%%
  %108 citations counted in INSPIRE as of 24 Oct 2017
  %\cite{Ambjorn:1995rg}
%\bibitem{Ambjorn:1995rg}
  J.~Ambjorn, J.~Jurkiewicz and Y.~Watabiki,
  %``On the fractal structure of two-dimensional quantum gravity,''
  Nucl.\ Phys.\ B {\bf 454} (1995) 313,
 % doi:10.1016/0550-3213(95)00468-8
  [hep-lat/9507014].
  %%CITATION = doi:10.1016/0550-3213(95)00468-8;%%
  %112 citations counted in INSPIRE as of 24 Oct 2017
  
  
  
  
   %\cite{Coleman:1988tj}
\bibitem{coleman}
  S.~R.~Coleman,
  %``Why There Is Nothing Rather Than Something: A Theory of the Cosmological Constant,''
  Nucl.\ Phys.\ B {\bf 310} (1988) 643.
 % doi:10.1016/0550-3213(88)90097-1
  %%CITATION = doi:10.1016/0550-3213(88)90097-1;%%
  %719 citations counted in INSPIRE as of 20 Feb 2017
  

%\bibitem{guth} 
 % A.~Guth, Phys.\ Rev.\ D {\bf 23},\ 347 (1981).


%\cite{Sotiriou:2008rp}
\bibitem{sotiriou}
  T.~P.~Sotiriou and V.~Faraoni,
  %``f(R) Theories Of Gravity,''
  Rev.\ Mod.\ Phys.\  {\bf 82} (2010) 451,
  [arXiv:0805.1726 [gr-qc]].
  %%CITATION = doi:10.1103/RevModPhys.82.451;%%
  %1830 citations counted in INSPIRE as of 30 Aug 2017





  
  
 
  
   


   
 
\end{thebibliography}
\end{document}